\documentclass[sigconf]{acmart}
\usepackage{xcolor}

\usepackage{multirow}
\usepackage{colortbl}
\usepackage{tabularx}
\usepackage{algorithm2e}
\usepackage{algpseudocode}
\newcolumntype{Y}{>{\centering\arraybackslash}X}
\usepackage{fontawesome}
\definecolor{MistyRose}{rgb}{0.99, 0.91, 0.95}
\definecolor{myyellow}{rgb}{1,0.96,0.56}
\AtBeginDocument{%
  \providecommand\BibTeX{{%
    \normalfont B\kern-0.5em{\scshape i\kern-0.25em b}\kern-0.8em\TeX}}}

\setcopyright{none}
\renewcommand\footnotetextcopyrightpermission[1]{}
\RestyleAlgo{ruled}

\begin{document}



\title{Trinity: Syncretizing Multi-/Long-tail/Long-term Interests All in One}

\author{Jing Yan, Liu Jiang, Jianfei Cui, Zhichen Zhao*, Xingyan Bin, Feng Zhang and Zuotao Liu}
\email{{yanjing.rec,jiangliu.reco,cuijianfei.ies,binxingyan,feng.zhang,michael.liu}@bytedance.com, *zhaozc10@gmail.com}
\affiliation{%
  \institution{ByteDance}
  \city{Beijing}
  \country{China}
}

\newcommand\blfootnote[1]{%
  \begingroup
  \renewcommand\thefootnote{}\footnote{#1}%
  \addtocounter{footnote}{-1}%
  \endgroup
}
\newcommand{\jiangl}[1]{\textcolor{red}{#1}}
\newcommand{\jianfei}[1]{\textcolor{orange}{#1}}

\renewcommand{\shortauthors}{Yan and Jiang, et al.}

\begin{abstract}
  Interest modeling in recommender system has been a constant topic for improving user experience, and typical interest modeling tasks (e.g. multi-interest, long-tail interest and long-term interest) have been investigated in many existing works.
  However, most of them only consider one interest in isolation, while neglecting their interrelationships.
  In this paper, we argue that these tasks suffer from a common ``interest amnesia" problem, and a solution exists to mitigate it simultaneously. We figure that long-term cues can be the cornerstone since they reveal multi-interest and clarify long-tail interest. Inspired by the observation, we propose a novel and unified framework in the retrieval stage, ``Trinity", to 
  solve interest amnesia problem and improve multiple interest modeling tasks. We construct a real-time clustering system that enables us to project items into enumerable clusters, and calculate statistical interest histograms over these clusters.
  Based on these histograms, Trinity recognizes underdelivered themes and remains stable when facing emerging hot topics. 
  Trinity is more appropriate for large-scale industry scenarios because of its modest computational overheads.
  Its derived retrievers have been deployed on the recommender system of Douyin, significantly improving user experience and retention.
  We believe that such practical experience can be well generalized to other scenarios.
\end{abstract}





\maketitle

\section{Introduction}


\begin{figure}[t]
  \centering
  \includegraphics[width=1.0\linewidth]{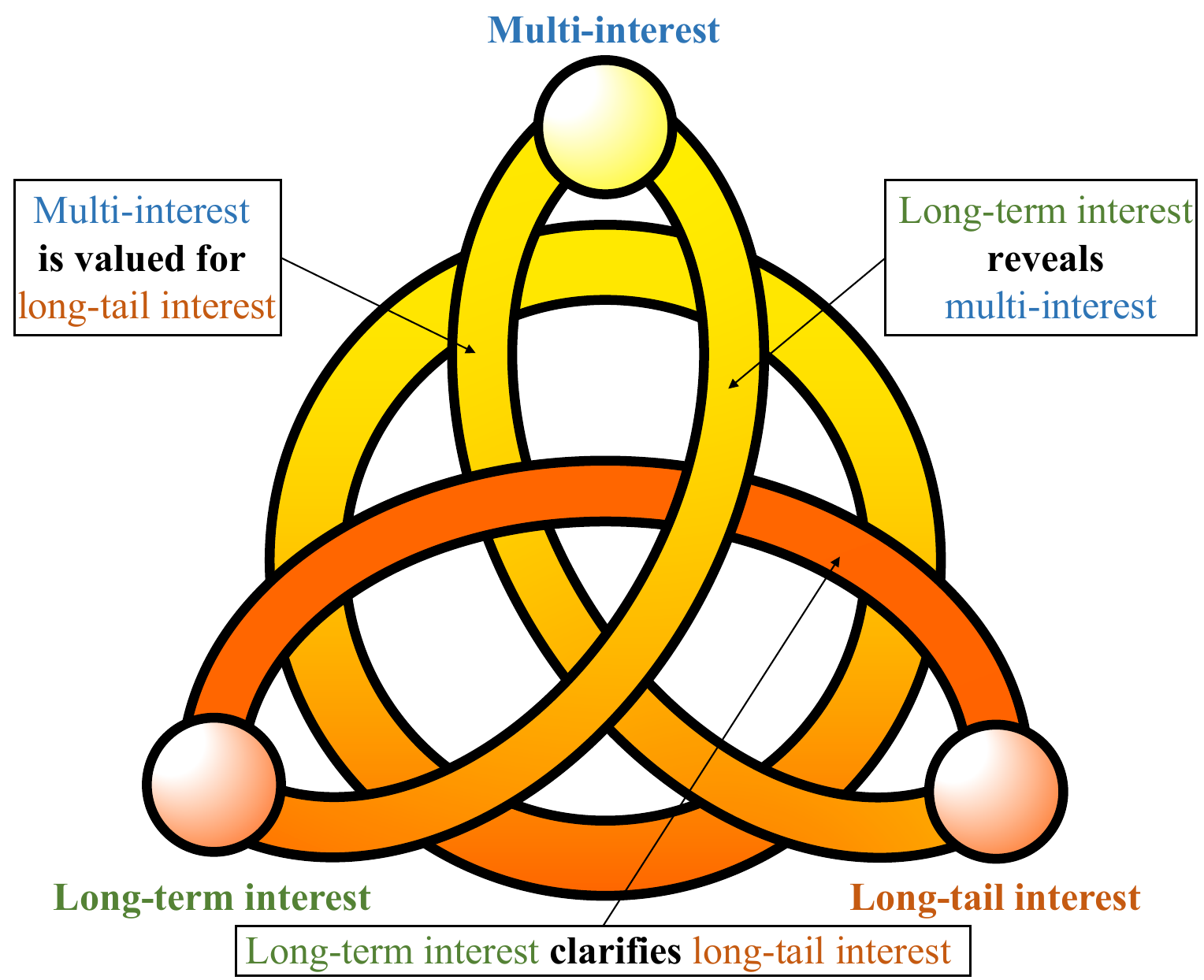}
  \caption{The interest reciprocity relationship and inspiration of Trinity. Multi-/long-term/long-tail interests are mutually dependent and reinforcing. See text for detailed examples.}
  \label{fig:trinity}
\end{figure}

\begin{figure*}[t]
  \centering
  \includegraphics[width=0.8\linewidth]{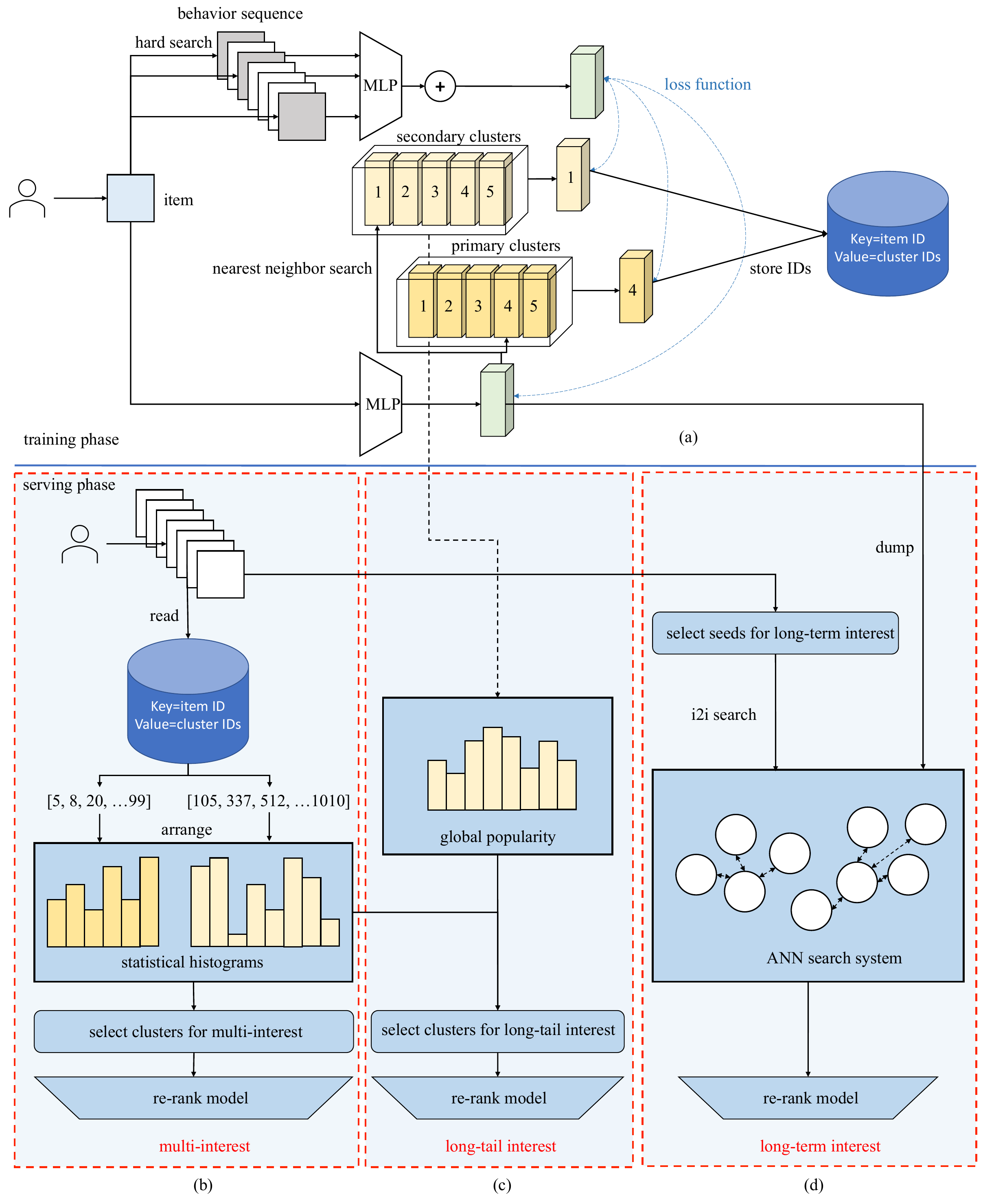}
  \caption{The framework of the proposed Trinity: (a) The training procedure. (b) By projecting user's long-term behavior sequence into histograms and customized strategy, multi-interest can be captured. (c) By comparing user's behavior with global cluster popularity, interests on long-tail themes can be clarified. (d) Time-agnostic embeddings produced in the training phase are applied for an i2i search retrieval.}
  \label{fig:framework}
\end{figure*}

\begin{figure}[t]
  \centering
  \includegraphics[width=1.0\linewidth]{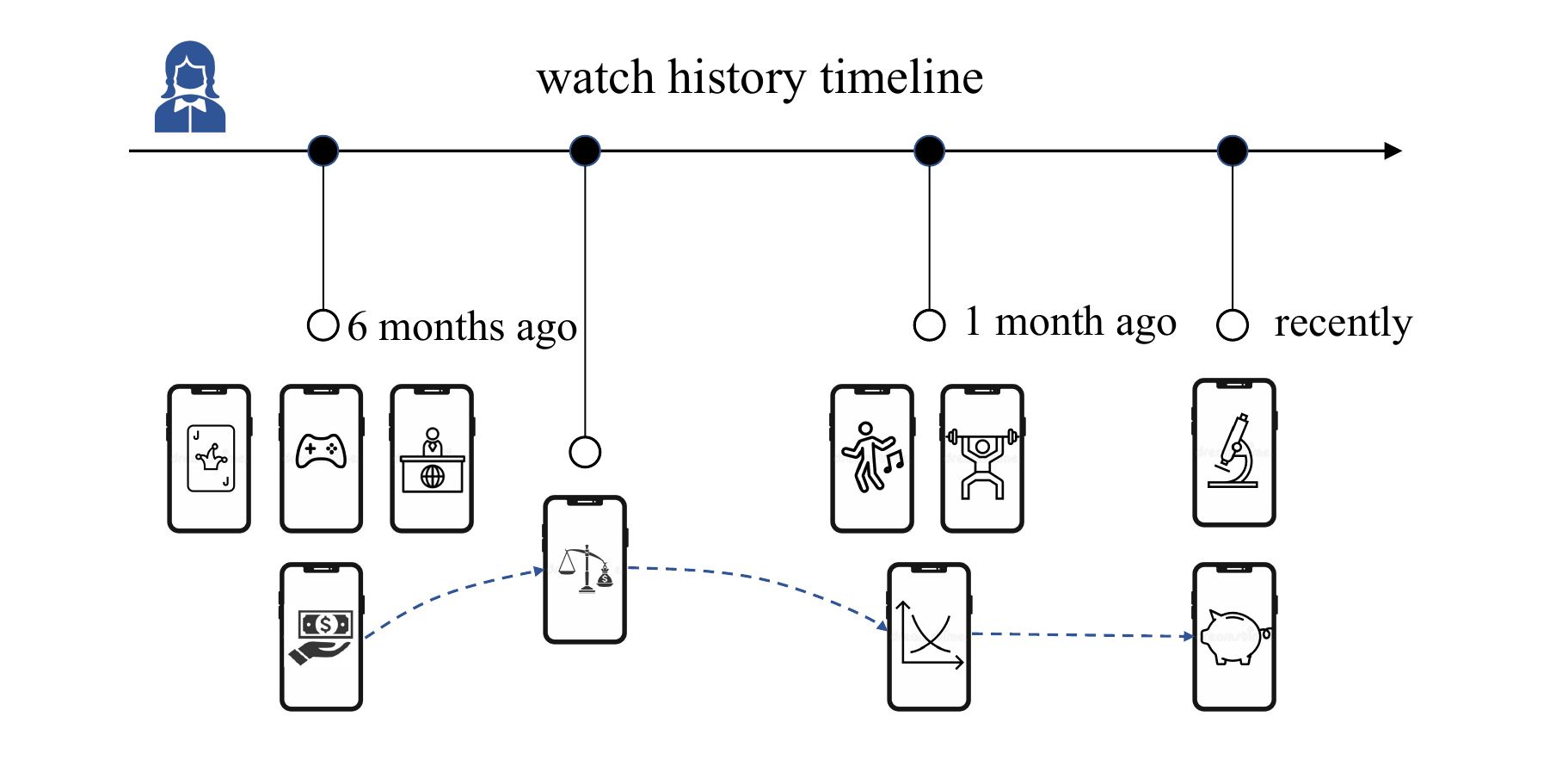}
  \caption{Watch history of a sampled user which reveals her short-term interests like games/talk shows and long-term interest (connected by dashed lines) on economics.}
  \label{fig:example}
\end{figure}

Large-scale industrial Recommender Systems (RS) confront numerous requests daily, and are supposed to capture each user's interest.
With the growth of daily active users (DAU) as well as their time spending on the platform, interest modeling becomes more and more challenging and diversified.
The emerging problems can be categorized as (1) multi-interest, which means users may be interested in many themes and topics (In Douyin, most users have more than $10$ interested topics based on our proposed clustering system).
(2) long-tail interest, referring to unpopular themes that only attract niche audience.
(3) long-term interest, which mainly describes themes that users are willing to watch consistently, rather than recent hot topics or emerging trend. 
For example, in Fig.\ref{fig:example} one user watched many game and talk show videos about six months ago. In the last month, she was obsessed with dancing and fitness themes. Nowadays she watches some research progress, while she has been following economic news for years. Here all topics mentioned above can be defined as multi-interest. Games/talk shows/dancing/fitness are long-term interest topics that may occasionally be ``forgotten" by our system, but still attract her if delivered. Research progress is a typical long-tail theme that is hardly appealing for everyone, but can be enjoyed by considerable users (that is why we also focus on long-tail interest). In this case, research progress has a chance to become multi-interest.

The key idea of this paper is that we believe the multi-/long-tail/long-term interests are mutually dependent and reinforcing. Therefore, we aim to improve these interest modeling tasks simultaneously.
As illustrated in Fig.\ref{fig:trinity}, we have some insightful observations. First, a user's recent behavior sequence can be easily dominated by several hot topics, so we need long-term cues to identify the rest topics of multi-interest. Then, in a completed industry system, popular themes are already well-delivered and multi-interest we are considering essentially refers to global long-tail themes. 
For example, in Douyin most users accept news and snapshots while we have to match calligraphy and its audience.
Finally long-tail interest also relies on long-term cues since whether one is truly interested in a certain topic can only be confirmed by long observations.
Inspired by these observations, we consider improving multiple interest modeling tasks simultaneously.

However, Modeling these interests is critical but quite hard: in current system we employ online learning models to rank candidates, which are trained by streaming paradigm and naturally tend to newcoming samples. 
So they may ``forget" some topics of interest due to occasional misses of the corresponding training samples, leading to the ``interest amnesia" problem, and worsening the delivery. Existing methods focus on modifying online learning framework, and introduce additional predictive heads and features that record users' past behavior. However, such methods pay excessive computational overheads and they only consider one of interest modeling tasks while neglecting their mutual relationship. 

In this paper,
we figure that the interest amnesia problem is essentially caused by online learning framework itself, and modifying it may be ineffective. On the contrary, we propose a statistic-based framework named ``Trinity". Trinity comprehends long-term cues first and summarizes user interest into statistical histograms, by which we can easily distinguish multi-/long-tail/long-term topics. Then, 
the underdelivered topics are selected by customized strategies to improve user experience. Since Trinity is modeled based on long-term statistical information, it can be impregnable to represent multiple interests while withstanding interest amnesia problem. 
Trinity is composed of three individual retrievers: Trinity-M for multi-interest, Trinity-LT for long-tail interest and Trinity-L for long-term interest, as demonstrated in Sec.\ref{sec:implement}.
To our best knowledge, Trinity is the first work that syncretizes multiple interest modeling into a unified framework and innovatively introduces long sequence cues (length $\ge 1000$) into retrieval stage.

Compared with existing methods which focus on one of interest modeling tasks, Trinity has some remarkable advantages: (1) Based on statistical histogram, each interested cluster is clearly evaluated and will not be forgotten or neglected. (2) Since items are assigned to indices exclusively, the computational overheads with interested topic growth are linearly increased, which outperforms existing methods. (3) Based on its clustering system, the whole delivery procedure is more explainable and manageable. We could work out more advanced strategies to reach better user experience. 
Trinity excellently benefits industrial recommender system, and we have launched the mentioned retrievers in Douyin and Douyin Lite.

In summary, the contributions of this paper are: 
\begin{enumerate}
\item We reveal the relationship among multi-/long-tail/long-term interests, and demonstrate their mutual relationship.
\item We propose the novel and unified framework of Trinity, which improves multiple interest modeling tasks simultaneously.
\item We demonstrate detailed implementation of Trinity with affordable computational overheads, which advances large-scale industrial recommender system.
\end{enumerate}

\section{Related Work}

In this section we summarize existing works into multi-interest, long-tail interest and long-term interest fields, and discuss how our work is different from them.

Recently, one of the popular methods considering multi-interest modeling is MIND~\cite{mind}. It proposes multiple user representations (in this paper they are also denoted as ``heads") to retrieve candidates from multiple fields, and has been the mainstream approach in industry applications. ComiRec~\cite{comirec} further considers diversity for improving system performance. MVKE~\cite{mvke} uses sub-expert-networks called ``virtual kernels" to capture user preference on various tasks and topics, and combine them to output comprehensive results. Some works consider user interest from different aspects, e.g. Jiang~\cite{whataspect} et al. explore multi-scale time effects on user interests. ~\cite{dynamimemory,sparseinterest} introduce similar ideas into sequence recommendation scenarios. Xie et al.~\cite{rethinkmultiinterest} improve negative mining and routing algorithms to address training issues in multi-interest learning. Since these methods train multiple user representations to cover various user preference, we summarize them as ``Multi-U" class methods. All the methods above focus on modifying existing online learning framework, while our proposed work relies on long-term statistics.

Long-tail items lack of user feedbacks or collaborative data so there are many methods focusing on multi-modal or graph methods~\cite{graph1,graph2,graph3} to improve the representations and prediction.
Huang and Wu~\cite{huang_wu} extract product profiles and similar products in a topic model.
Kumar et al.~\cite{harness} realize long-tail recommendation using few shot learning, instead.
Note in this paper, we are not focusing exactly on long-tail items. Long-tail interest mainly refers to niche topics but their items are unnecessary to be obscure.

In the area of long-term interest, Pi et al.~\cite{longseq} directly improve long sequence modeling by
decoupling the most resource-consuming part from the entire model to a separate module. They update user representations by trigger events rather than requests and support thousands-length sequence.
The most well-known method of capturing long-term interest is Search-based Interest Model (SIM~\cite{sim}). It searches past topic-relevant behavior sequence to supplementarily measure user preference, but is mostly adopted in ranking models. 
Our approach avoids directing involving long sequence in models, which may be not well scalable. Instead, we exploit clusters and general interest distributions to mine long-term cues.

Our work employs VQ-VAE~\cite{vqvae} in the training procedure to assign items to indices, and construct a hierarchical clustering system meanwhile. It allocates cluster embeddings and permits items to find their nearest neighbors. By that we can project untold items into enumerable clusters, and cope with user behavior sequence.
Its varieties~\cite{wsdm22,kdd23} have been widely used in information retrieval. 

\section{The Basic Trinity Framework}\label{sec:trinity}

In this section, we first demonstrate the key idea of Trinity, and introduce its basic training framework. The details of improving multi-/long-tail/long-term interests with Trinity framework are organized in Sec.\ref{sec:implement}.

As illustrated in Fig.\ref{fig:trinity}, we consider multiple interests simultaneously. Different from existing work, we argue that modeling interests are not isolated tasks, on the contrary, these interests are mutually dependent and reinforcing. 
 So taking the whole situation into consideration greatly benefits the system.
We have three perspectives:

1) Long-term interest reveals multi-interest. Users' short-term interest can be dominated by emerging trends and unpredictable hot topics, resulting in homogeneous items in their behavior sequences. By involving longer behavior sequences, we can figure out which topics form multi-interest
and obtain more comprehensive user preference.

2) Multi-interest is valued for long-tail interest. 
Multi-interest offers a wide collection of content themes, in which hot topics and niche themes can be both captured.
However, items of hot topics (e.g. news, games) have been well estimated and delivered by existing retrievals and ranking stages in a completed recommender system.  When we are thinking of multi-interest, we are actually focusing on some uncommon themes, such as science and poems.

3) Long-term interest clarifies long-tail interest. Long-tail themes attract only a narrow audience, while many plays are just made by casual interest. To find out the real audience who consistently consume these themes we need to backtrack long-term behavior records.

For now, the interrelationships of interests are coherent, and inspire us to model them simultaneously. But it is still hard to implement it in the online learning framework. Existing works have been struggling to solve the interest amnesia problem. However, 
online learning models are forced to fit current samples so they inevitably tend to newcoming samples and popular items (occupy the most impressions). If samples of some topics become infrequent, the corresponding interests will be gradually forgotten. We conclude this phenomenon as ``interest amnesia problem", which needs enormous labour to be mitigated in online learning framework. 

Instead of modifying the online learning framework, in this paper, we try to employ another instrument:
\textbf{statistics}. Statistical cues extracted on long-term behavior can clearly reveal every interested topics, and remains stable when facing emerging trends. So it effectively mitigates interest amnesia problem and distinguishes multi-/long-tail/long-term interests.
Specifically, we setup an interest clustering system. Then, users' past behaviors are projected into the system, resulting in long-term behavior histograms. Finally, we can implement customized delivery strategies from the histograms. As long as we search long-term behaviors, long-term interest modeling is improved, and we improve the others as following:

(1) Sort the histogram descendingly by behavior count, choose the top-k clusters and disperse them, ensuring that items of middle clusters are not dominated by head clusters. This provides comprehensive results and improves multi-interest.

(2) For each global long-tail cluster, if the current user frequently consume items from this theme, we enhance its delivery. This matches niche themes with their audience, and thus improves long-tail interest.

Considering the implementation, we need the following instruments: a clustering system which can be constructed by embeddings and evenly measures long-term and recent behaviors, a storage system by which we can map behaviors into clusters and thus form users' histograms in real time, and a series of delivery strategies.

In Fig.\ref{fig:framework}(a) we illustrate the basic framework of Trinity. The training phase is mainly composed of a SIM (Search-based Interest Model~\cite{sim}) head and a VQ-VAE~\cite{vqvae} structure. 
For each item, we firstly search its SIM sequence, and extract embeddings for both (denoted as green cubes) as $\mathbf{x}$ and $[\mathbf{b}_1,\mathbf{b}_2,\dots \mathbf{b}_i]$, for $i\le N_b$. Behavior sequence embeddings are mean-pooled to form user-side representation as $\mathbf{b}=\sum_i \mathbf{b}_i / {N_b}$.
Meanwhile we prepare primary and secondary learnable clusters $c^1_j$, $c^2_k$ where $j$ in $1\dots J$, $k$ in $1\dots K$ (denoted as yellow cubes), and allocate their embeddings $\mathbf{e}_j^1$ and $\mathbf{e}_k^2$. $J$ and $K$ are set to be 128 and 1024.
We search the top-1 nearest neighbor clusters for the current item (e.g. ID=4 and 1 in Fig.\ref{fig:framework}) as $\mathbf{c}_{\hat{j}}^1$ and $\mathbf{c}_{\hat{k}}^2$ and apply Binary Cross Entropy (BCE) loss functions for ($\mathbf{b}$, $\mathbf{x}$), ($\mathbf{b}$, $\mathbf{e}_{\hat{j}}^1$) and ($\mathbf{b}$, $\mathbf{e}_{\hat{k}}^2$) pairs:

\begin{equation}
\begin{aligned}
L = \sum_p\sum_{\mathbf{A}=\mathbf{x},\mathbf{e}_{\hat{j}}^1,\mathbf{e}_{\hat{k}}^2} y_p {\rm log}(\sigma(\mathbf{b}_p^T\mathbf{A}_p)) + (1-y_p) {\rm log}(1-\sigma(\mathbf{b}_p^T\mathbf{A}_p)), 
\end{aligned}
\end{equation}
where subscript $p$ denotes sample index in a batch. here $y=1$ if the user finishes the video, or interacts (upvote/share/follow/comment) with it, or watches it for more than 10 seconds. Moreover, we add some random negative samples to obtain more discriminative embeddings. Note the assignment relationship between items and clusters are determined by training progress in real time by nearest neighbor search, and we dump such information into a key-value storage (key=item ID and value=[primary cluster ID, secondary cluster ID]).
Our implementation follows the same design in VQ-VAE where gradients for clusters are directly cast on items while cluster embeddings are updated by moving average of their belonging item embeddings.
Here we use a hierarchical clustering system of two levels. The fine-grained secondary level is designed to select items while the coarse-grained primary level avoids
repetitive topics and acts as dispersion (see Sec.\ref{sec:trinity-MI} for detailed example).

Since item embeddings are trained by a SIM head, which involves behaviors record in a very wide time range. Recent items and very ``old" items are optimized simultaneously and evenly measured. 
This affords us a collaboration-based, time-variant and temporal-unbiased clustering system (which can not be achieved by multi-modal features). 
More importantly, infinite items are discretized into enumerable cluster set, which makes cluster-level statistics possible. 


\section{Interest Modeling in Large-scale Recommender System}\label{sec:implement}
In the section above, we have demonstrated the training procedure. In this section, we introduce specific and practical strategies to solve interest amnesia problem and improve multi-/long-tail/long-term interest modeling respectively. Note in this paper, we only focus on the retrieval stage. The reason is two-fold:
on one hand, retrievers determine candidates for ranking models. If we lose topics of interest at this stage, ranking models can do nothing to improve interest modeling.
On the other hand, retrievers always employ Approximate Neighbor Nearest (ANN) search system for computational efficiency. In accurate calculation scenarios where ranking models evaluate each candidate exhaustively, we can boost some topics by dispersion to ``remember" them.
However, we can hardly employ the same idea because approximate calculation discards some candidates at the very beginning. 

\subsection{Multi-interest}\label{sec:trinity-MI}
The key idea of improving multi-interest modeling is selecting clusters with moderate but significant intensity in user's statistical histograms, which have already been consumed by users but are easy to be forgotten by models. 
Given behavior counts on each cluster, well-delivered, underdelivered and uninterested topics are clearly revealed.

As illustrated in Fig.\ref{fig:framework}(b), in the serving stage, for each request we first extract user's long-term behavior sequence (length=2500). Note the sequence only stores item IDs that $playtime \ge 10s$ or have been finished or interacted (upvote/follow/share/comment). Then, the corresponding cluster IDs are read from the key-value storage mentioned in Sec.\ref{sec:trinity} in real time, and are organized into primary and secondary histograms $\mathbf{h}^1=[h^1_1,h^1_2,\dots h^1_J]$ and $\mathbf{h}^2=[h^2_1,h^2_2,\dots h^1_K]$ where each bar refers to behavior count of the $j$/$k$-th cluster.

By simply sorting the histograms descendingly, user's interest distribution is clearly revealed. For example, if a sorted histogram is $[50,20,20,4,2,0,0,0]$ and the corresponding indices are $[10,33,100,91,62,21,5,83]$, cluster whose ID=10 is the user's major interest (we define ``major interest" clusters which own the strongest response). However, there also exists significant response at index=33 and 100, which constitutes ``multi-interest" we concentrate in this paper (major interests are also part of multi-interest, but they always have been well-delivered by existing retrievers). Clusters of 91 and 62 are treated as exploring interests that may become user's multi-interest or global long-tail interest. Other clusters are ignored.

Inspired by such observation, we introduce a strategy-based retriever to improve multi-interest modeling. First, we read user's past behavior cluster IDs: ${(c^1_i,c^2_i)}_{i=1}^L$ where $L\le 2500$. In each pair, secondary cluster ID is attached as primary cluster's child node. It results in structural data like $c^1_j[h^1_j] \Rightarrow c^2_{jk}[h^2_{jk}]$ where ``$\Rightarrow$" denotes one-to-many mapping relationship. Note in the training stage items individually search their nearest clusters so primary and secondary clusters are only statistically related. There may exist same secondary clusters under different primary clusters and we need dual subscripts in $h^2_{jk}$. In the following procedures, primary clusters are used as dispersion method, and we only leverage secondary clusters to deliver items.

First, we select primary clusters where $h_j^1$ is greater than $T_p=30$ and each of whose secondary cluster has $h_{jk}^2$ greater than $T_s=10$. Then we randomly choose one secondary cluster for each primary cluster. If there are not enough clusters, we select the secondary cluster from primary clusters above with the greatest $h_{jk}^2$. If there is still not enough clusters, we collect secondary clusters one-by-one from largest to smallest. In each step above, we remove duplicated clusters.
The algorithm is summarized in Alg.\ref{alg:multiinterest}, which forms an additional retriever named ``Trinity-M" by adding a re-rank model to evaluate the candidates of selected clusters.

Some middle clusters can be dominated by major interested topics since model bias (model believes most items of cluster A outperforms ones of cluster B), so in an unitary two-tower retriever, they cannot obtain enough impressions. However, in Trinity-M we only select one secondary cluster in each primary cluster. This acts like dispersion and we can collect complementary cluster set, which are not forgotten. By ``awakening" these topics, users get part of their interest back, and feel more satisfied with our system.
In Sec.\ref{sec:cc} we show the complementariness and comprehensiveness of Trinity-M, which leads to significantly improved user experience as well as more diversified system.
Note we involve long sequence cues in Trinity framework by projecting behaviors into clusters, while avoiding straight employing in online learning models (retrievers have to manage billions of candidates, 
we can hardly use SIM~\cite{sim} like ranking models). To our best knowledge, Trinity is the first work that successfully employs long sequence cues ($\ge 1000$) in the retrieval stage.

\begin{algorithm}
\caption{Trinity retriever for multi-interest}\label{alg:multiinterest}
\KwData{Structural data $c^1_j[h^1_j] \Rightarrow c^2_{jk}[h^2_{jk}] $, threshold $T_p=30$, $T_s=10$, $N_{M}=10$}
\KwResult{Secondary cluster set $R_{M}$}
$R_{M} \gets \emptyset$\;
$p \gets 0$\;
\ForEach{$c^1_j$}{
  \If{$h^1_j\ge T_p$ and $\forall c^2_{jk}$, $h^2_{jk}\ge T_s$}{
    randomly choose one secondary cluster $c^2_{j\hat{k}}$\;
    \If{$c^2_{j\hat{k}} \notin R_{M}$}{
        $R_{M} \gets R_{M}\cup \{c^2_{j\hat{k}}\}$\;
    }
  }
}
\If{$|R_{M}|\ge N_{M}$}{
    $R_{M} \gets \{randomchoose(R_{M},N_{M})\}$\;
    \Return $R_{M}$\;
}
\ForEach{$c^1_j$}{
  \If{$h^1_j\ge T_p$}{
    $c^2_{j\hat{k}} \gets argmax(h^2_{jk})$\ , where $c^2_{jk} \notin R_M$\;
    $R_{M} \gets R_{M}\cup \{c^2_{j\hat{k}}\}$\;
  }
}
\If{$|R_{M}|\ge N_{M}$}{
    $R_{M} \gets \{randomchoose(R_{M},N_{M})\}$\;
    \Return $R_{M}$\;
}
sort all $c^2$ by $h^2$ from largest to smallest\;
\While{$|R_{M}|<N_{M}$}{
    \If{$c^2_{p} \notin R_{M}$}{
        $R_{M} \gets R_{M}\cup \{c^2_{p}\}$\;
        $p\gets p+1$\;
    } 
}
\Return $R_{M}$\;
\end{algorithm}

\begin{table}[]
\centering
\begin{tabularx}{\linewidth}{cYY}
\hline
Aspect     & Multi-U  & Trinity-M \\ \hline
Flexibility &  \faTimesCircle & \faCheckCircleO \\ \hline
\multirow{ 2}{*}{Efficiency}   & deteriorates with  & consistently keeps  \\ 
 & many heads & linear increasing \\ \hline
\multirow{ 2}{*}{Explainability}  & semantically  & semantically  \\ 
 & ambiguous & meaningful \\ \hline
\multirow{ 2}{*}{Extensibility} & has not & has been achieved   \\ 
 & been validated & in this paper \\ \hline
Fungibility& feasible & only complementary \\ \hline
\end{tabularx}
\caption{Comparison between Multi-U class methods and Trinity-M.}
\label{tab:multiuid}
\end{table}

\begin{figure}[t]
  \centering
  \includegraphics[width=1.0\linewidth]{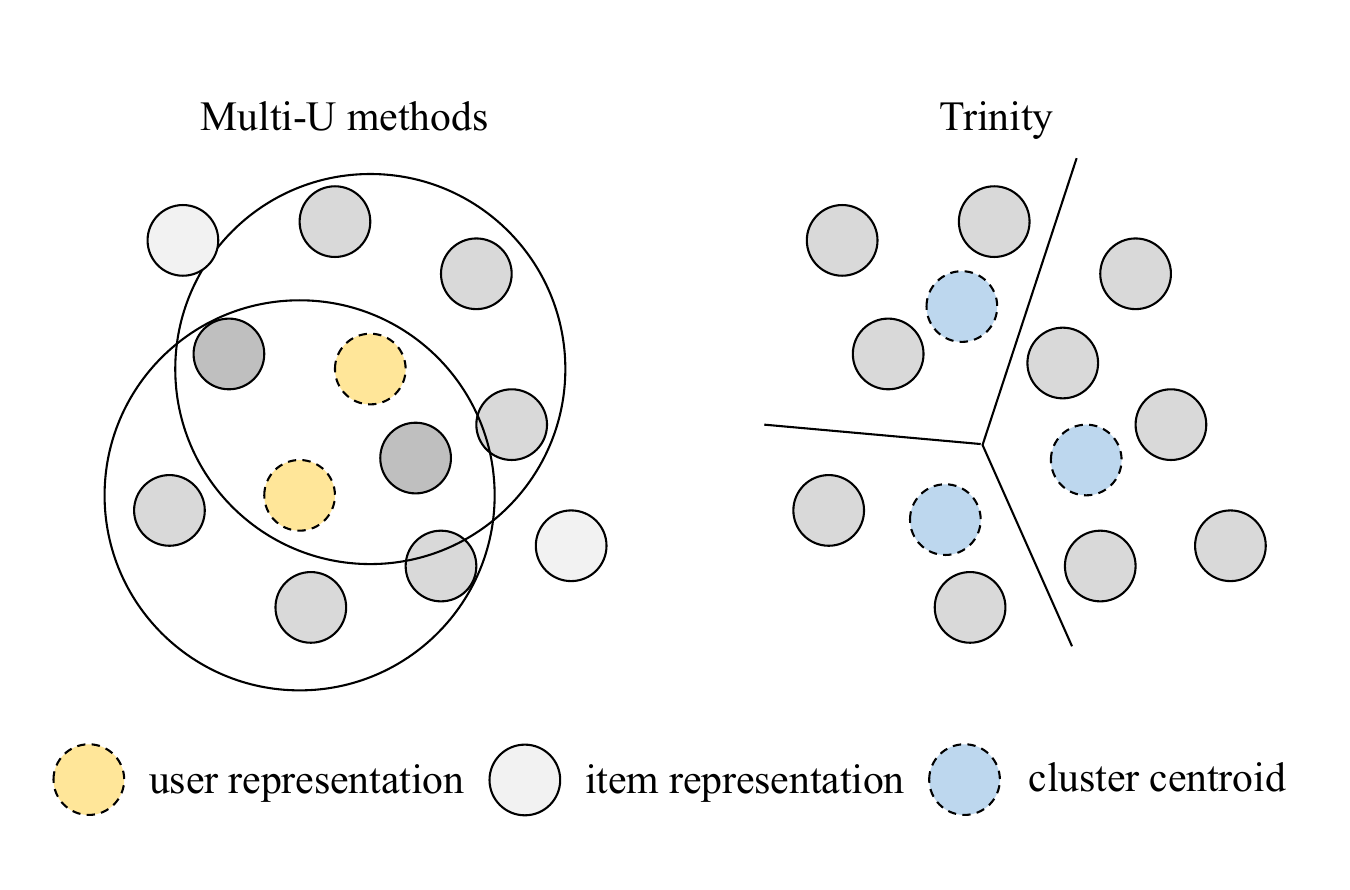}
  \caption{In Multi-U methods, each user representation (yellow dashed circle) searches candidates individually, so some items may be retrieved (solid hollow circles) more than once (deeper gray refers to more retrievals). With the head count increasing, redundant computational overheads grow heavily. However, in Trinity items are assigned exclusively, each item can be retrieved by at most once.}
  \label{fig:multiu}
\end{figure}

We compare advantages and disadvantages of Trinity-M with the ``Multi-U" class methods in Table.\ref{tab:multiuid}. Here we use ``Multi-U" to summarize most existing works that train several user representations (heads) and search multiple candidates results for capturing multi-interest. The most well-known Multi-U methods are MIND~\cite{mind}, ComiRec~\cite{comirec} and MVKE~\cite{mvke}. 
\begin{enumerate}
\item \textbf{Flexibility}. It measures whether we can adjust our interested topics flexibly. In Trinity-M the strategy is customized so extending considered topics, skipping head clusters and implementing topic sampling would be easy. However, Multi-U methods provide no effortless solution without training new models.

\item \textbf{Efficiency}. The efficiency describes how serving cost increases with more considered topics. For Multi-U methods, when we introduce an additional head, we execute once more u2i search. However, as illustrated in Fig.\ref{fig:multiu}, items can be redundantly
searched by multiple heads. So the more heads we have, the less unique candidates we can retrieved by adding heads.
On the contrary, in Trinity-M items are assigned to clusters deterministically and exclusively. When we add additional clusters, they only bring supplementary candidates. Finding more interested clusters consistently costs linearly additional overheads.

\item \textbf{Explainability}. The explainability describes semantic meaning of specific heads/clusters. In Multi-U methods heads are restrained to learn different samples, however, we can hardly control the training procedure or appoint specific topics to be optimized. In trinity-M, clusters themselves are semantically meaningful (see Sec.\ref{sec:clustering}), and we know what topics we are considering.
\item \textbf{Extensibility}. The extensibility measures whether the method can be extended to other scenarios. Multi-U methods propose seldom solution for long-tail or long-term interest modeling tasks. While Trinity simultaneously improves these interest modeling problems, as demonstrated below. 
\item \textbf{Fungibility}. The fungibility refers to whether we can use the proposed method to replace existing retrievers. For this aspect, Multi-U methods are alternative because their multiple heads easily cover candidates of a single u2i model. However, since Trinity is modeled based on long-term statistic, emerging items/topics are hard to be prominent in time. So we use Trinity-M as an additional retriever and only consider its supplementary function.

\end{enumerate}

\subsection{Long-tail Interest}\label{sec:trinity-LT}
For long-tail interest, our idea is to compare personal interest distribution 
against global interest distribution based on clusters. First, we need to automatically recognize/define global long-tail clusters. Then, we check if there exists significant response in user's statistical histogram on these clusters. Clusters meeting this criteria are collected to form an additional retriever.

Inspired by~\cite{samplingbias}, we employ the similar streaming frequency estimation, but apply it on clusters rather than items. Specifically, cluster $c_k$ (for long-tail interest we only consider secondary clusters) is mapped by a hash function $\mathcal{H}$, and when it occurs in the training streaming, its occurrence interval is calculated by
\begin{equation}
\begin{aligned}
B[\mathcal{H}(c_k)]\leftarrow (1-\alpha)\cdot B[\mathcal{H}(c_k)] + \alpha \cdot (t-A[\mathcal{H}(c_k)]), 
\end{aligned}
\end{equation}
where $A[\mathcal{H}(c_k)]$ records the last occurrence timestamp of the cluster. $t-A[\mathcal{H}(c_k)]$ denotes the latest occurrence interval and $B[\mathcal{H}(c)]$ is its global moving average. This is illustrated in Fig.\ref{fig:framework}(c).

In general, we can recognize global long-tail themes from clusters of large occurrence intervals. However, large occurrence interval clusters are not only composed by long-tail themes. There are also some ``juxtapose" clusters which represent hot topics, but seize few items because of training occasionality. We remove clusters that have less than $T_i=3$ items and effectively avoid juxtapose clusters. In our implementation, we define top 600 clusters (cover about 22\% impressions) ranked by occurrence intervals as long-tail clusters without juxtapose clusters.

After defining long-tail clusters, we check user's statistical histogram. If a cluster has more than $T_l=3$ behavior count, it is regarded as ``convinced" response on long-tail interest, and is involved.
We introduce a sampler to select $N_{LT}=20$ clusters where each cluster has probability of
\begin{equation}
\begin{aligned}
Pr(c_k)=\frac{(\beta + h_k)^{\alpha}}{\sum_{q=1}^K (\beta+h_q)^{\alpha}}
\end{aligned}
\end{equation}
to be sampled. The hyper parameter $\alpha$ adjusts the distribution (for example, $\alpha=0$ forms uniform distribution) and we use $\beta$ as a smooth term. In our implementation we set $\alpha=0.75$ and $\beta=0.1$. The designed sampler outperforms a uniformly random sampler on user experience as well as diversity since items belonging to top clusters are more probable to be consumed (see Sec.\ref{sec:online}).
Note duplicated clusters with Trinity-M have been removed at first.

The algorithm above is summarized in Alg.\ref{alg:longtailinterest}. After collecting
secondary clusters, we feed their items into a re-rank model to obtain retrieval results. This retriever is denoted as Trinity-LT.

\begin{algorithm}
\caption{Trinity retriever for long-tail interest}\label{alg:longtailinterest}
\KwData{$c^2_{k}[h^2_{k}] $, global occurrence intervals ${B[\mathcal{H}(c)]}$, threshold $T_i=3$, $T_l=3$, $N_C=600$, $N_{LT}=20$}
\KwResult{Secondary cluster set $R_{LT}$}
$R_{LT} \gets \emptyset$\;
$C \gets \emptyset$\;

\ForEach{$c_k$}{
  \If{$c_k$ has equal or more than $T_i$ items}{
    $C \gets C\cup \{c_{k}\}$\;
  }
}

sort clusters ${c}_k \in C$ by $B[\mathcal{H}(c_k)]$, from largest to smallest\;
choose top $N_C$ clusters as $C$\;

\ForEach{$c^2_k$}{
  \If{$c^2_k\in C$ and $h^2_k\ge T_l$ and $c^2_k \notin R_{LT}$}{
    $R_{LT} \gets R_{LT}\cup \{c^2_{k}\}$\;
  }
}
$R_{LT} \gets \{sampler(R_{LT},N_{LT})\}$\;
\Return $R_{LT}$\;
\end{algorithm}

\subsection{Long-term Interest}
As illustrated in Sec.\ref{sec:trinity}, SIM head optimizes user's long past behavior items and current items simultaneously. So we believe the generated item embeddings represent long-term cues, and construct an i2i retriever to capture user's long-term interest.

As illustrated in Fig.\ref{fig:framework}(d), we train a pre-rank model to select seeds from user's past behavior sequence. To capture comprehensive interest, any $playtime\ge 10s$, finish or interaction (upvote/share/comment/follow) feedback is treated as positive label. The model follows conventional two-tower architecture, and involves random negative samples.

After ranking user's behavior items (length $\le$ 2500), we project them into Trinity clustering system, and disperse by limiting that candidates of the same cluster can not exceed $T_c$. Then we randomly select $N_L$ from top $N_s$ ranked items as seeds, and search items with similarity measured by Trinity embeddings. Similar with Trinity-M and Trinity-LT, searched items are fed into the re-rank model, and form the retriever of Trinity-L. Note long-term cues have already been leveraged by Trinity-M and Trinity-LT, and we introduce Trinity-L as a lightweighted and supplementary extension.

\begin{figure*}[h]
  \centering
  \includegraphics[width=0.9\linewidth]{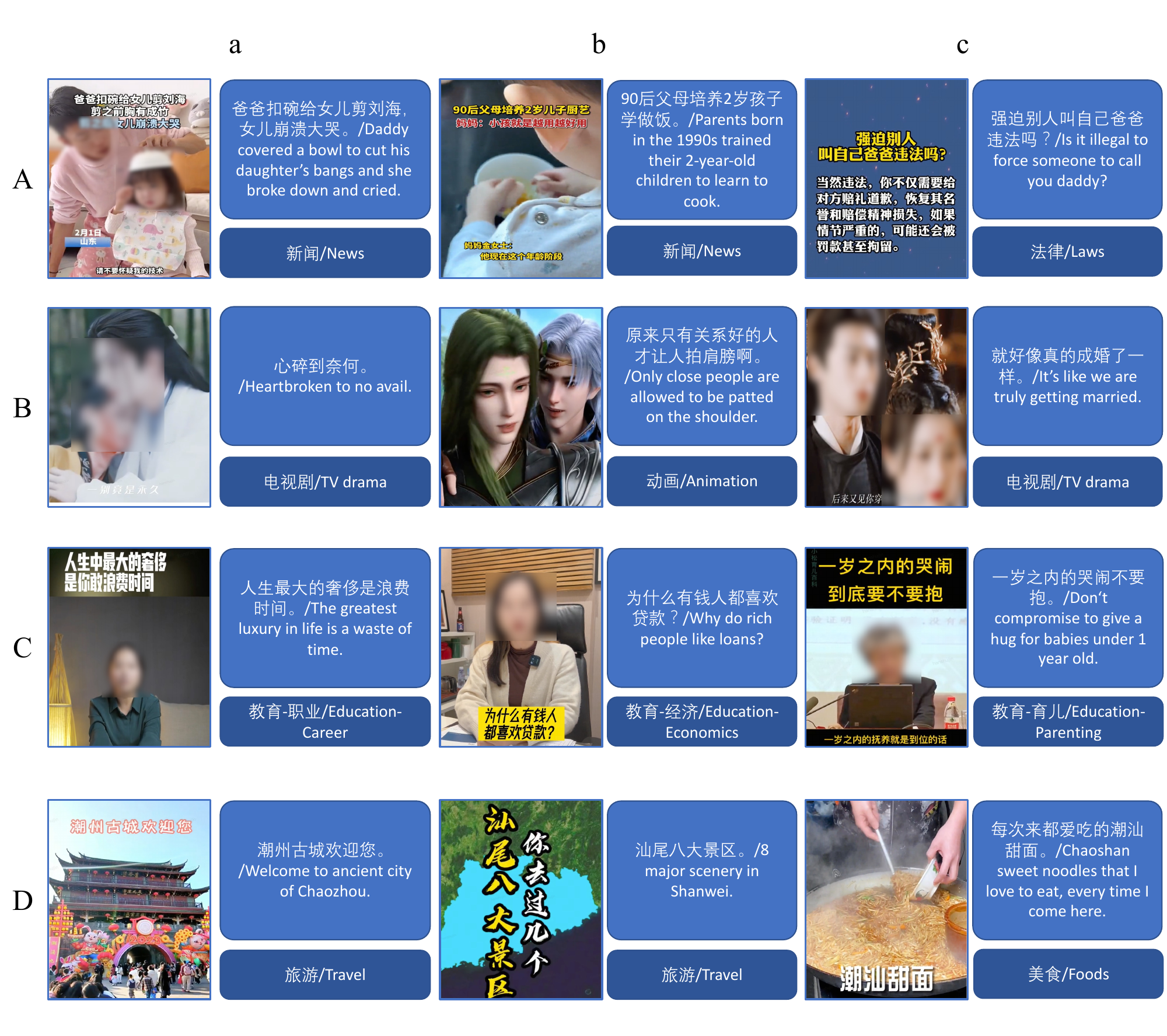}
  \caption{The visualization of items belonging to Trinity clusters. Items of the same row share the same secondary cluster ID.}
  \label{fig:vis}
\end{figure*}

\subsection{Implement Details}\label{sec:details}
The proposed three retrievers select clusters and aggregate their belonging items, while remaining 10K-100K candidates, which is still too large for the following ranking stages. So we employ re-rank models to shrink candidates to about 1000. The re-rank models follow two-tower architecture, but precisely predict ranks without ANN in the serving phase.

The target of re-rank models is to predict video play time. Specifically, samples that are watched within 2 seconds are negative. For the rest positive samples, we use their play time (clipped to be at most 5 minutes) to weight the in-batch Softmax loss~\cite{samplingbias}. In fact such a model has been deployed as a major retriever called ``stay-time retriever", and we just reuse it as re-rank models for additional retrievers.

\section{Experiments}
In this section, we present the performance of Trinity. First, we show visualized analysis on its clustering system, demonstrate some interesting properties of the proposed model.
Then, we show online performance of the additional retrievers on our large-scale industry scenarios. 
We also analyze them in detail to figure how they improve interest modeling tasks as well as user experience.

\subsection{Clustering System of Trinity}\label{sec:clustering}
In Fig.\ref{fig:vis} we visualize some items belonging to Trinity clusters. Items of the same row share the same secondary cluster ID, and we use (cluster ID, item ID) to refer them.
The most important difference between Trinity clustering system and an intuitive clustering system from human perspective (denoted as label system), is its clustering manners.
For example, in the first row, all items are about parent-child relationship rather than parenting, which is uncommon from human perspective. These items come from multiple tags such as news and laws, even the last example (A,c) talks about imaginary parent-child relationship that actually doesn't happen. The similar phenomenon can be observed in the cluster of B, where items are all about handsome men
despite he is a real person or an animation model. The theme of a cluster focuses on an unconventional topic, and covers multiple tag aspects.

The third row shows a much common taxonomy example where three items are all about education, and each of them explains knowledge of a sub-topic (career, economics and parenting).
The most surprising example comes from the last row. The items show scenic spots and foods in three places, while these places are in the same province.
So we can conclude another property of Trinity clustering: its taxonomy is varying with general topics. In the education topic, it categorizes items by subjects, while in the travel topic, it divides items into geographical bins.

\subsection{Online Experiments}\label{sec:online}
\begin{table*}
\centering
\begin{tabularx}{\textwidth}{c|YYYY|YYYY}
\toprule \hline
 &
  \multicolumn{4}{c|}{Douyin} &
  \multicolumn{4}{c}{Douyin Lite} \\ \cline{2-9} 
\multirow{-2}{*}{Retrievers} &
  \multicolumn{1}{c}{Watch Time} &
  \multicolumn{1}{c}{AAD} &
  \multicolumn{1}{c}{AAH} &
  \multicolumn{1}{c|}{AT} &
  \multicolumn{1}{c}{Watch Time} &
  \multicolumn{1}{c}{AAD} &
  \multicolumn{1}{c}{AAH} &
  \multicolumn{1}{c}{AT} \\ \hline
MIND~\cite{mind} (6 heads*) &
  +0.093\% &
  +0.009\% &
  +0.033\% &
  -0.472\% &
  +0.176\% &
  - &
  +0.051\% &
  -0.270\% \\
Trinity-M &
  +0.118\% &
  +0.008\% &
  +0.046\% &
  +0.153\% &
  +0.178\% &
  +0.018\% &
  +0.078\% &
  +0.038\% \\ \hline
Trinity-LT &
  +0.069\% &
  - &
  +0.019\% &
  +0.546\% &
  +0.060\% &
  - &
  - &
  -
   \\
Trinity-LT (uniform sampling) &
  - &
  - &
  -  &
  +0.154\% &
  - &
  - &
  - &
  -
   \\ \hline
Trinity-L &
  +0.051\% &
  +0.009\% &
  +0.020\% &
  -0.080\% &
  +0.069\% &
  +0.014\% &
  +0.029\% &
  -0.081\%
   \\ \hline
 \bottomrule
\end{tabularx}
\caption{Results of online A/B experiments, measured by Watch Time, Average Active Days (AAD), Average Active Hours (AAH) and Average Tags (AT).}
\label{tab:online_results}
\end{table*} 

We conduct online A/B experiments and show the results of large-scale industrial recommender system on Douyin and Douyin Lite.
As mentioned above, we simply launch three additional retrievers: Trinity-M Trinity-LT and Trinity-L onto the whole system, and report the performance.

\noindent \textbf{Metrics}. The ultimate goal of recommendation algorithm is improving user experience, which is widely measured by Daily Active Users (DAU). However, DAU can hardly be observed in A/B experiments so we need alternative metrics. In this paper, we use Average Active Days (AAD) as a surrogate metric to describe user experience. When a user first arrives the platform, he/she is randomly grouped into control or experimental group. Then we count his/her active days during the experimental period, and calculate the average difference between the two groups.
For most applications, DAU is the most critical metric. However, in Douyin, there are many users who log in almost everyday, so AAD can not describe their experience. To tackle this problem, we propose an additional metric Average Active Hours (AAH) which calculates users' active hours similar with AAD. 
For multi-interest and long-tail interest modeling, diversity is expected to be improved. So we also involve Average Tags (AT) which counts how many tags are consumed over all items. 
In addition, we also employ Watch Time as an auxiliary metric. 

In Table.\ref{tab:online_results} we show the online performance, where only statistically significant metrics are listed. 
We have exhaustedly verified Multi-U methods, none of them obtains significant improvement within acceptable computational overheads. We report the results of MIND~\cite{mind} here with 6 serving heads and omit others for concise. Such cost far exceeds our budget and the method is infeasible. 
One of unexpected observations is that MIND diminishes diversity, which implies all these heads repetitively retrieve some hot topics.
On the contrary, Trinity-M significantly improve AAD and AAH on Douyin and Douyin Lite under limited overheads, and is successfully deployed.
We provide an ablation study on Trinity-LT where uniform sampling means using a uniformly distributed sampler to select clusters. The group of uniform sampler has no positive impact on user experience and less diversity improvement, which verifies that clusters with larger response of a user are more likely to be consumed.
Trinity-L also significantly improves AAD and AAH on Douyin and Douyin Lite. One interesting observation is that Trinity-L contributes much less AAH but reaches competitive AAD improvement compared with Trinity-M. This implies that multi-interest modeling mainly benefits high-active users (active more than 20 days in the past month). It consists with our expectation since high-active users are more likely to discover new interested topics.

\begin{figure}[h]
  \centering
  \includegraphics[width=1.0\linewidth]{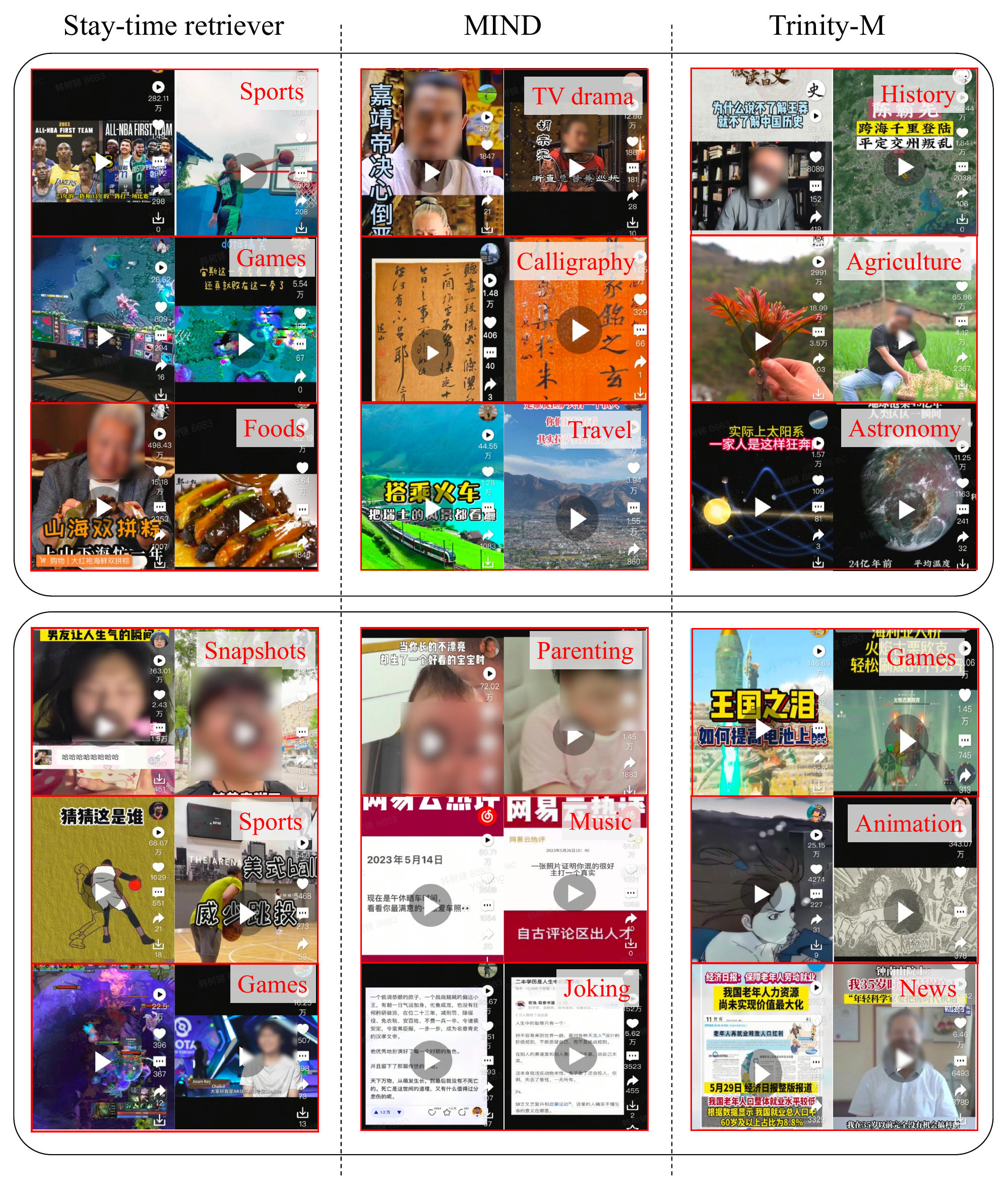}
  \caption{Top retrieved results of multiple retrievers. Two users are divided by boxes, and we only show supplementary topics of MIND and Trinity-M.}
  \label{fig:case}
\end{figure}

\subsection{The Comprehensiveness and Complementariness of Trinity-M}\label{sec:cc}
In Fig.\ref{fig:case} we visualize topics that can be retrieved by stay-time retriever, MIND and Trinity-M for 2 users. Here we show top-ranked results since only these candidates can be fed into the following ranking stages. Note MIND and Trinity-M can retrieve their left topics, but we omit them for concise visualization.

For the first user, stay-time retriever only provides popular topics such as sports and games. MIND retrieves several additional topics like TV drama and travel. Here calligraphy is a pretty positive case since it is a long-tail topic. Trinity-M provides more topics (history, astronomy and agriculture) in addition. Trinity-M's retrieved clusters are not only more comprehensive, but also tend to long-tail topics.

The example of the second user is much different, however. This user's interest mainly comes from popular topics such as games, parenting and news. Surprisingly, even hot topics can not be entirely retrieved by the unitary stay-time retriever. And Trinity-M provides complementary results of news and animation. Note the ``Games" in the last column is different from the one in the first column, since it is a sub-category describing console games.
In Douyin, most users have more than 10 interested secondary clusters measured by Trinity's clustering system, so only Trinity-M meets our demand of modeling multi-interest.

\subsection{What contents benefit from Trinity-LT?}\label{sec:distribution}
The motivation behind Trinity-LT is improving global long-tail themes delivery, and in our implementation we use taxonomy provided by Trinity itself. In this part we verify long-tail themes delivery based on label system instead, which mainly comes from human perspective. From our practical experience, when items estimation is improved, their long-term impressions obtain significant increase, vice versa. So we can simply check impression distribution change caused by Trinity-LT to validate its influence on long-tail interest modeling. 

\begin{table}[]
\centering
\begin{tabularx}{0.9\linewidth}{ccccc}
\hline
  Themes & Finance  & Laws & Photography & Career \\ \hline
 VV & +0.904\%  &  +0.478\% & +0.389\% & +0.230 \\ \hline
 \hline
  Themes & Movie  & Games & Variety & TV drama \\ \hline
 VV & -0.520\%  &  -0.349\% & -0.349\% & -0.322\%  \\ \hline
\end{tabularx}
\caption{Impression distribution change caused by Trinity-LT.}
\label{tab:trinity-LT}
\end{table}

In Table.\ref{tab:trinity-LT} we show the impression distribution change mentioned above where VV (Video View) refers to impressions of the specific theme. Some typical boosted themes are finance (+0.904\%), laws (+0.478\%), photography (+0.389\%) and career (+0.230\%). In our platform they are all recognized as niche topics.
Note the results are calculated across the whole market, in Trinity-LT these themes are boosted even more significantly.
On the contrary, hot topics occupy less impressions thus Trinity-LT matches niche themes with their audience and generally deboosts popular themes, which meets our expectation. 

\subsection{Seed Distribution of Trinity-L}
Here we analyze the properties of Trinity-L.
Since each item in user's 2500-length past behavior sequence can probably be selected as a seed, we expect that Trinity-L will retrieve earlier seeds. Compared with the existing i2i retriever whose seeds are selected by rules on recent behavior and employs conventional online learning model embeddings, Trinity-L has more seeds of 7-15 days ago (+67\%) and 15-30 days ago (+28\%). In the experiment Trinity-L delivers the earliest seeds for mid-active users (active more than 10 but less than 20 days in the past 30 days) and obtain significant AAD improvement on this group. It suggests that the refined long-term interest modeling leads to better user experience, 
and Trinity-L indeed captures long-term cues to improve user experience.
Note "refined" here needs Trinity embeddings since without them we cannot obtain significant AAD improvement even on long-term seeds.

\section{Conclusion}
In this paper we focus on multi-/long-tail/long-term interest modeling at retrieval stage. We figure that online learning framework hardly solves 
the widely existing interest amnesia problem, and propose a unified and statistic-based framework to tackle the problem: Trinity. Based on a collaborative and time-variant clustering system, Trinity
derives three individual retrievers (Trinity-M, Trinity-LT and Trinity-L) for the corresponding interest aspects. Trinity-M and Trinity-LT support customized strategies and 
retrieve targeted topics. While Trinity-L recovers some interested topics that have been forgotten. All of the retrievers have been deployed on Douyin and Douyin Lite, and have significantly improved user experience. We believe long-term statistic directly mitigates interest amnesia, and is tailored for large-scale industry systems.

\bibliographystyle{ACM-Reference-Format}
\bibliography{sample-base}










\end{document}